\documentclass[namedreferences,hyperref,optionalrh,solaromanenum]{spr-sola}
\usepackage{amssymb}
\usepackage{color}           % For color text: \color command

\usepackage{soul}
\usepackage{todonotes}
\usepackage{hhline}

\usepackage{graphicx,graphics,amsmath}
\usepackage{natbib}
\usepackage{bm,url}
\usepackage{times}
\usepackage{float}
\usepackage{arydshln}
\graphicspath{{./fig/}{./png/}}
\newcommand{\EQ}{\begin{equation}}
\newcommand{\EN}{\end{equation}}
\newcommand{\EQA}{\begin{eqnarray}}
\newcommand{\ENA}{\end{eqnarray}}

\newcommand{\del}{\mbox{\boldmath $\nabla$}}
\newcommand{\brac}[1]{\langle #1 \rangle}

\newcommand{\pd}{\partial}

\newcommand{\DIV}{\bm{\nabla} \cdot }
\newcommand{\bdot}{\mbox{\boldmath $\cdot$}}

\newcommand{\Th}{\overline{T}}
\newcommand{\Sh}{\overline{S}}
\newcommand{\Ph}{\overline{P}}
\newcommand{\rh}{\overline{\rho}}

\newcommand{\BB}{\bm{{B}}}

\newcommand{\UU}{\bm{{U}}}

\newcommand{\oom}{\mbox{\boldmath $\Omega_0$}}

\newcommand{\cross}{\mbox{\boldmath $\times$}}
\newcommand{\curl}{\mbox{\boldmath $\nabla \times$}}

\newcommand{\vv}{\bm{{{v}}}}
\newcommand{\grav}{{\bf g}}

\newcommand{\mps}{m~s$^{-1}$}

\def\mc{meridional circulation}
\def\mf{meridional flow}
\def\dr{differential rotation}

\def\bl{Babcock--Leighton}

\def\cz{convection zone}

\def\to{torsional oscillation}
\def\Rs{R_{s}}
\newcommand{\rhat}{\hat{\mbox{\boldmath $r$}} {}}
\newcommand{\thetahat}{\hat{\mbox{\boldmath $\theta$}} {}}
\newcommand{\phihat}{\hat{\mbox{\boldmath $\phi$}} {}}

\newcommand{\Eq}[1]{Equation~(\ref{#1})}
\newcommand{\Eqs}[2]{equations~(\ref{#1}) and~(\ref{#2})}

\newcommand{\Fig}[1]{Figure~\ref{#1}}
\chardef\us=`\_

\begin{document}

\begin{frontmatter}
\title{Solar cycle variations of meridional flow and differential rotation in a non-kinematic dynamo model} %with imposed large-scale flows}

\author[addressref={aff1}, corref,email={karak.phy@iitbhu.ac.in}]{\inits{B.B.}\fnm{Bidya~Binay}~\snm{Karak}\orcid{0000-0002-8883-3562}}
\author[addressref={aff2},email={mark.miesch@noaa.gov}]{\inits{M.}\fnm{Mark}~\snm{Miesch}\orcid{0000-0003-1976-0811}}
\address[id=aff1]{Department of Physics, Indian Institute of Technology (Banaras Hindu University), Varanasi 221005, India}
\address[id=aff2]{Cooperative Institute for Research in Environmental Sciences, University of Colorado, Boulder, CO, US
}

\runningauthor{Karak \& Miesch}
\runningtitle{Solar cycle variations of meridional flow and differential rotation}

\begin{abstract}
Flux transport dynamo models have emerged as the leading paradigm for explaining the observed solar cycle. However, most of these models are kinematic and axisymmetric. We develop a non-kinematic dynamo model by specifying large-scale flows, such as the differential rotation and meridional flow. Thus, our model does not require the $\Lambda$ effect or other parameterisations of turbulent transport to generate the large-scale flows; it only solves the fluctuating flows and thermodynamics.
Our model is inherently 3D by construction, although it is currently operating in axisymmetric mode because the source for poloidal field generation (Babcock-Leighton $\alpha$) is assumed to be axisymmetric.  Our model reproduces solar-cycle variations in both the meridional flow and the differential rotation — specifically, the torsional oscillation in some agreement with observations. The model naturally exhibits some variability in the solar cycle, even without incorporating stochastic fluctuations, offering a new methodology for modelling the non-kinematic solar dynamo in 3D. 
\end{abstract}
%\keywords{MHD, Solar Dynamo, Solar Cycle, Differential Rotation, Meridional Circulation, Torsional oscillation, Magnetic field}
\end{frontmatter}
%-------------------------------------------------

\section{Introduction}
\label{sec:int}
An important axisymmetric large-scale flow of the Sun is the \dr. In earlier days, its existence was inferred by tracking successive images of the sunspot groups.
It was known that the equatorial region of the Sun rotates faster than higher latitudes 
\citep{C63}. From surface measurements, \citet{HL80} discovered a small periodic variation in rotation with the sunspot cycle, known as \to.
Later, these results were confirmed by helioseismology, which provides \dr\ in the entire solar \cz\ (CZ) \citep{Thomp03}. Additionally, helioseismic measurements showed two branches of \to, namely, an equator-ward migrating branch at low latitudes and a poleward branch at high latitudes with a relatively higher amplitude \citep{H09}.

The second important axisymmetric large-scale flow in the CZ is the \mc. Near
the surface, it is poleward with an amplitude of about 15~\mps. The structure of the flow and its amplitude inside the CZ are debated at present, 
although there must be a return
flow at some depth \citep{RA15, Gizon20}. 
Interestingly, unlike \dr\ for which a temporal variation of only a few per cent in terms of \to,
was detected, the meridional flow shows a considerable temporal variation.
Notably, an apparent solar cycle variation of the meridional flow of about $55\%$ is observed---becoming weaker
during the solar maximum and stronger during the solar minimum \citep{Gonz06, HR10}.
These two large-scale flows, along with the non-axisymmetric convective flow in the rotating solar 
CZ produce a dynamo action, which is manifested in the observed solar cycle \citep{Miesch05, Cha20}.

To model these large-scale flows and the magnetic field in the Sun, nonlinear global magnetohydrodynamics (MHD) simulations have been carried out
first by \citet{Gilman83} and \citet{Glatz84, Glatz85}. These exploratory simulations, although pivotal, have failed to reproduce many basic features of the large-scale flows and the magnetic field.
Even after 40 years of these works, global convection simulations that use significantly enhanced computing power still struggle to reproduce observed flows, particularly the solar differential rotation \citep[see also][and the references therein]{KMB18b, Hotta22, kapyla_ISSIRev}. 

However, some progress has been made in modelling large-scale flows using hydrodynamic (HD) mean-field models \citep[e.g.,][]{KR95, KS01, rempel05, HY11, KOdr11}. 
These models require parameterisation of turbulent transport, which may, however, involve free parameter(s). 
Most of these models do not include the back-reaction of the dynamo-generated magnetic field.

Flux transport dynamos, in particular, have been successful in reproducing some basic features
of the solar magnetic field at least in some parameter regimes \citep[see reviews by][]{Kar14a, Cha20, Hazra_rev21, Kar23, Hazra23}. 
However, these models are mostly kinematic; therefore, the large-scale flows are not evolved but are instead imposed. Moreover, most of these models are 2D (axisymmetric) with a few exceptions; see \cite{MD14, MT16, KM17, KM18, HM18, WYM19, Kumar19, TM24} for 3D and \citet{LCC15, LC17} for coupled 2D$\times$2D kinematic dynamo models.
In these models, the generation of the poloidal field from the decay and dispersal of tilted bipolar magnetic regions, the Babcock--Leighton process, is explicitly modelled.

In the literature, nonkinematic dynamo models have been presented at different levels.
In some models, the feedback of the large-scale magnetic field is included through a quenching term in the imposed meridional flow \citep{KC12, Passos12}, while in other models \citep{Kuk99, Kit99, Covas00, CCC09}, the feedback of the magnetic field is included in the equation of differential rotation
through the mean-field Lorentz force and/or, $\Lambda$ effect (in addition to the usual quenching in $\alpha$ effect).
The motivations of the latter studies were to explore the solar cycle variability and/or torsional oscillation.
In other models, magnetic feedback was included in the full momentum equation mainly to explore dynamo saturation \cite[e.g.,][]{Br92, Moss95}.
\citet{HC17} solved the perturbation of the mean meridional flow by including the Lorentz force of the toroidal field obtained from a flux transport dynamo model to study how the meridional flow changes in response to the solar cycle.
In \citet{Remp06} and \citet{HY11}, the mean-field equations for differential rotation and meridional circulation were solved
in conjunction with the equations for the \bl\ type flux transport dynamo model. In this model, the magnetic feedback was included via the mean-field Lorentz force and the quenching of transport coefficients (turbulent viscosity and heat conductivity); see also \citet{BC22} for an extension of this model to 3D. 
Despite different formulation levels, what is common to all these models, except \citet{BC22}, is that they are all axisymmetric. Therefore, the non-axisymmetric magnetic field and tachocline instabilities that could be excited through interactions with the flow and the magnetic field are not possible in these models.

In this study, we shall build a non-kinematic \bl\ dynamo model in which we specify the background mean flows but solve for their perturbations. Hence, we do not have to model the large-scale flows by specifying the $\Lambda$ effect. 
Our approach is similar to those in \citet{HC17} and \citet{SM19}, but we shall consider the full MHD, and the perturbed flows will act back on the magnetic field.  
In this model, we study the interaction between the flow
and the magnetic field and particularly the variation of meridional flow and torsional oscillation.
We also explore the effect of the perturbed flow produced by the Lorentz force on the dynamo and on solar cycle variability. 
The model is intrinsically 3D by construction, as it is built on the Rayleigh code; however, we present the results of the axisymmetric case because the Babcock--Leighton $\alpha$ term is considered axisymmetric. 

\section{Model}
\label{sec:mod}
We build our model using the open-source code Rayleigh \citep{Featherstone24};  see also \citet{FH16, matsui16} for some applications using this code.
In Rayleigh, MHD equations are prescribed using the anelastic approximation 
and by linearising the thermodynamic variables about their spherically symmetric, 
time-independent reference state with density $\rh$, pressure $\Ph$, temperature $\Th$, and
specific entropy $\Sh$. Fluctuations about this thermodynamic state are denoted
without overbars. Thus, the equations that Rayleigh solves are as follows \citep[see][ for details]{FH16}.
\begin{equation}\label{eq:continuity}
\DIV \left(\rh \vv \right) = 0
\end{equation}
%--- Momentum equation ---
\begin{eqnarray} \label{eq:mom}
 %  \begin{split}
\frac{\pd \vv}{\pd t} = - \left(\vv \bdot \del\right) \vv - 2 \oom \cross \vv - \del\left(\frac{P}{\rh}\right) - \frac{S}{c_p} \grav  \\
+ \frac{1}{\rh}  {\bm \nabla \cdot} {\bm {\mathcal  D}}
%+ \frac{1}{\rh} \DIV \DD
+ \frac{1}{4\pi\rh} \left(\curl \BB\right) \cross \BB \equiv {\bf R}_v
 \nonumber
 %  \end{split}
\end{eqnarray}
%---Entropy equation---
%\begin{eqnarray} \label{eq:S}
\begin{equation} \label{eq:S}
%  \begin{split}
\rh \Th \frac{\pd S}{\pd t} = - \rh \Th \vv \bdot \del S + \DIV \left(\rh \Th \kappa \del S\right) + Q_i + Q_o + Q_v 
     \equiv \rh \Th R_S
% \nonumber
%  \end{split}
%\end{eqnarray}
\end{equation}

%--- Induction equation ---
\begin{equation}\label{eq:indy}
\frac{\pd \BB}{\pd t} = \curl \left(\vv \cross \BB\right) - \curl\left(\eta \curl \BB\right) \equiv {\bf R}_B
\end{equation}

Here $\bm\Omega_0=(\cos\theta,-\sin\theta,0)\Omega_0$ is the angular velocity vector, $\grav$ is the gravitational acceleration, $c_p$ is the specific heat at constant pressure,
$\mathcal{D}_{ij} = -2 \rh \nu \left[e_{ij} - \frac{1}{3}(\del \cdot \vec{v})\delta_{ij} \right]$,
where $e{_{ij}}$ is the strain rate tensor, $\nu$ is the kinematic viscosity, and $\delta_{ij}$
is the Kronecker delta, 
$\kappa$ is the thermal diffusivity, $Q_i$ is the internal heating, which includes the divergence of the radiative flux, $Q_o$ is the ohmic heating,
$Q_v$ is viscous heating, and $\eta$ is the magnetic diffusivity.

Now we impose a flow by replacing $\vv$ in Equations (\ref{eq:continuity})--(\ref{eq:indy}) 
with $\vv + {\UU}$.  Here ${\UU}$ is the steady (time independent) axisymmetric \dr\ and \mc\ 
so that,
\begin{equation}
{\UU} \equiv {\UU}(r,\theta) = U_r \rhat + U_\theta \thetahat + U_\phi \phihat. ~~~.
\end{equation}
The corresponding thermodynamic variables are replaced as follows 
\begin{eqnarray}
P \rightarrow & P + P_U(r,\theta) \\
S \rightarrow & S + S_U(r,\theta) \\
T \rightarrow & T + T_U(r,\theta) \\
\rho \rightarrow & \rho + \rho_U(r,\theta)  ~~~.
\end{eqnarray}

Also, we assume that
\begin{equation}
\DIV \left(\rh {\UU}\right) = 0
\end{equation}
so equation (\ref{eq:continuity}) is unchanged.

%On subsititing these variables into \Eq{eq:mom}, we 
The next step is to substitute these changes into equations (\ref{eq:mom})--(\ref{eq:indy}).  
However, before we do that, let us first define the pressure such that it is in a dynamical balance with the flow:
\begin{equation}\label{eq:P0}
\del\left(\frac{P_U}{\rh}\right) = - \left({\UU} \bdot \del\right) {\UU} 
- 2 \oom \cross {\UU} - \frac{S_U}{c_p} \grav + \frac{1}{\rh} \DIV \bm {\mathcal D}_U + {\bf R}_c,
\end{equation}
where $\bm {\mathcal  D}_U$ is the viscous diffusion tensor computed with ${\UU}$ and ${\bf R}_c$ are hypothetical Reynolds and Maxwell stresses that serve to maintain the mean flows $\UU$.
These lie outside the scope of the model.  In fact, there is no need to compute what they are and we just need to postulate that they are there and that they are responsible for the time-independent mean flows.

For $S_U$, we assume the thermal wind balance:
\begin{equation}
 2 \Omega_0 \frac{\partial U_\phi}{\partial z} = \frac{g}{r c_p} \frac{\pd S_U}{\pd \theta}, ~~~
\end{equation}
(with $z=r \cos\theta$)
which gives
\begin{equation}
\frac{\pd S_U}{\pd \theta} = \frac{2 r \Omega_0 c_p}{g} ~ \left(\cos\theta \frac{\pd U_\phi}{\pd r} - \frac{\sin\theta}{r} \frac{\pd U_\phi}{\pd \theta}\right) \equiv  \mathcal{L}. ~~~
\end{equation}
Applying a convenient differential operator to this gives:
\begin{equation}
\frac{1}{\sin\theta} \frac{\pd}{\pd \theta} \left(\sin \theta \frac{\pd S_U}{\pd \theta} \right) = 
\frac{1}{\sin\theta} \frac{\pd}{\pd \theta} \left(\sin^2 \theta \left[\frac{\mathcal{L}}{\sin\theta}\right]\right).   ~~~
\end{equation}

Hence, we can compute $S_U$ from $U_\phi$ and, in principle, we could compute $P_U$ from \Eq{eq:P0}, but we do not really need to.  
It is sufficient to simply assume that the balance exists.  The linearised equations of state then imply:
\begin{equation}
\frac{\rho_U}{\rh} = \frac{P_U}{\gamma \Ph} - \frac{S_U}{c_p}
\end{equation}
and
\begin{equation}
T_U = \Th ~ \left(\frac{P_U}{\Ph} - \frac{\rho_U}{\rh}\right)  ~~~.
\end{equation}
Again, we do not have to compute these -- this is just to show that everything is consistent.

Similarly, we can assume that the following balance exists in the thermal energy equation:
\begin{equation}\label{eq:Sbalance} 
- \rh \Th {\UU} \bdot \del S_U + \DIV \left(\rh \Th \kappa \del S_U\right) + Q_U + Q_c + Q_i = 0 
\end{equation}
where $Q_U$ is the viscous heating associated with ${\UU}$ and $Q_c$ is a hypothetical energy flux divergence due to convective motions that are beyond the scope of the model.  Note that $Q_i$ is included in the balance because the convective energy flux should carry out the heat buildup that arises from the convergence of the radiative flux.

So, with these assumptions, the substitutions into equations (\ref{eq:mom})--(\ref{eq:indy}) 
become:
\begin{equation}\label{eq:momnew}
\frac{\pd \vv}{\pd t} = {\bf R}_v - \left({\UU} \bdot \del\right) \vv 
- \left(\vv \bdot \del\right) {\UU}  
\end{equation}

\begin{equation}\label{eq:Snew}
\frac{\pd S}{\pd t} = R_S - {\UU} \bdot \del S - \vv \bdot \del S_U - \frac{Q_i}{\rh \Th}  
\end{equation}

\begin{equation}\label{eq:indynew}
\frac{\pd \BB}{\pd t} = {\bf R}_B  + \curl \left({\UU} \cross \BB\right) + {\bm \nabla} \times \left( S_{\rm BL} \phihat \right) ~~~.
\end{equation}

For the background axisymmetric mean-flow ($\UU$) we obtain the same profile as given in 
the kinematic dynamo benchmark publication \citep{Jouve08}. Therefore, the analytical profile 
and parameters to obtain $U_r$ and $U_\theta$ (meridional flow) and $U_\phi$ 
(obtained from differential rotation) 
are the same as given in case SC$^\prime$ of \cite{Jouve08}.

As our model does not capture the toroidal flux emergence, formation of tilted active regions and their subsequent decay to the generation of the poloidal field on the solar surface \citep[however see:][for such approach]{YM13, Kumar19}, here we adopt the traditional prescription of generation of poloidal field in the flux transport dynamo models \citep{CSD95, NC01}. For this, we have  
%For the generation of the poloidal field from the toroidal field in our model, 
%we follow the usual prescription
%of the flux transport dynamo model and 
added the following source term 
%${\bm \nabla} \times \left[ S_{\rm BL}(r,\theta,t) \phihat \right]$ 
%\[
${\bm \nabla} \times \left( S_{\rm BL} \phihat \right)$
%\]
in \Eq{eq:indynew}. Here, $S_{\rm BL}$ captures the generation of poloidal field through the decay of tilted sunspots near the solar surface at low latitudes, mimicking the Babcock--Leighton process. Following \citet{Jouve08}, the profile of $S_{\rm BL}$ is given by 
\begin{eqnarray} \label{eq:al}
    S_{\rm BL} (r,\theta,t) = \alpha_0 \frac{1}{2} \left[ 1 + {\rm erf}\left(\frac{r - r_1}{d1}\right) \right] \left[ 1 - {\rm erf}\left(\frac{r - R_\odot}{d_1}\right) \right] \\
\times    \left[1 + \left(\frac{B_\phi(r_c,\theta,t)}{B_0} \right)^2  \right]^{-1}
    \cos\theta \sin\theta B_\phi (r_c,\theta,t), 
\end{eqnarray}
where $\alpha_0 = 15$ cm~s$^{-1}$, $r_1 = 0.95 R_\odot$, $d_1 = 0.01 R_\odot$, and $r_c = 0.7 R_\odot$.

The magnetic diffusivity $\eta$, which is taken equal to $\nu$
is obtained from the following equation \citep{Jouve08}.
%{\bf\blue{ (SC$^\prime$ case) of the benchmark model
% \cite{Jouve08} (their Eqs. 18 and 14, respectively).}}
\begin{equation}
\eta(r) = \eta_c + \frac{1}{2}\left(\eta_t-\eta_c \right) \left[1+{\rm erf}\left(\frac{r-r_c}{d}\right) \right],  
\end{equation}
with $\eta_c = 10^{9}$~cm$^{2}$ s$^{-1}$,  $\eta_t = 3 \times 10^{10}$~cm$^{2}$ s$^{-1}$, and $d = 0.02R_\odot$.

We impose impenetrable stress-free boundary conditions for the perturbed velocity
at both radial boundaries. The entropy perturbation is set to zero at these boundaries.
For the magnetic field, the radial field on the solar surface 
and perfect conductor at the bottom boundary are taken.

%%%%%%%%%%%%%%%%%%%%% Result %%%%%%%%%%%%%%%%%%%%%%%%%%%%%%%%%%%%%%%%%%%%%%%%

\section{Results from the kinematic dynamo case}
\label{sec:kin}
We observe that with the initial condition $\vv = S = P = 0$,
and $Q_i = 0$ (the imposed flows carry no net heat flux),
all extra terms in \Eqs{eq:momnew}{eq:Snew}
vanish. However, the advection term due to the mean flow in \Eq{eq:indynew} 
does not vanish  
 (this together with $S_{\rm BL}$ will indeed sustain the dynamo\footnote{We have explicitly checked that without this source $S_{\rm BL}$, even in the non-kinematic case, the initial seed magnetic fields in our simulations decay due to diffusion.}). 
 %{\bf\blue{Sec. 3 of \citet{Cha10}}}).
%(initial seed field $\BB \neq 0$---this is what will sustain the dynamo). 
%Thus, initially, the only non-zero term in $R_v$ is the Lorentz force, and the only non-zero term in $R_S$ is the ohmic heating $R_o$. 
We set both the Lorentz force in $R_v$ and $R_o$ to zero in the kinematic case.
%Essentially, in the kinematic case, we make both the Lorentz force and $R_o$ zero.
That is, we evolve only the magnetic induction \Eq{eq:indy}
with flow replaced by the imposed mean flow.

\begin{figure}
\centering
\includegraphics[scale=0.35]{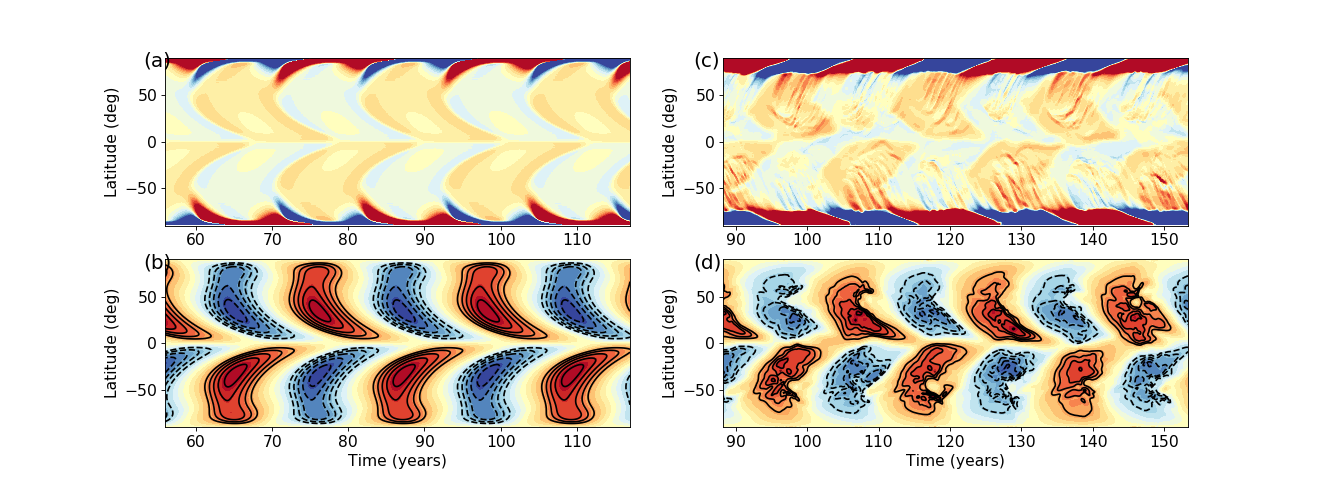}
\caption{The time-latitude (butterfly) diagram of the surface radial field $B_r(R_\odot,\theta,t)$ 
(top) and the toroidal field at the base of the CZ 
$B_\phi(0.71R_\odot,\theta,t)$ 
(bottom). Left and right panels are obtained from the kinematic and non-kinematic models, respectively. 
}
\label{fig:bfly}
\end{figure}

The time-latitude (popular butterfly) diagrams of $B_r$ and $B_\phi$ from this model are shown 
in the left panels of \Fig{fig:bfly}.
We note that the nonlinear quenching used in the Babcock--Leighton type $\alpha$ effect
halts the dynamo growth rate and results in this stable solution. 
We find that these plots resemble the basic features of the observed magnetic cycle (polarity reversal, equatorward migration, and 11-year periodicity). Our magnetic cycles are also similar to the results obtained in other kinematic flux transport dynamo models with similar parameters, in particular, to those presented in Figure~14 of \citet{Jouve08} and in Figure~4 of \citet{MT16}). 
The cycle period is longer in our case because we use a lower diffusivity and $\alpha_0$ than those used in those models.

\section{Results from non-kinematic dynamo model}
\label{sec:nonkin}
Now we present the results of the non-kinematic model in which we evolve 
the perturbed flow and entropy
equations along with the magnetic field equations (Equations \ref{eq:momnew}--\ref{eq:indynew}).

\subsection{Magnetic field and cycles}
\label{sec:mag}
\begin{figure*}
\includegraphics[scale=0.28]{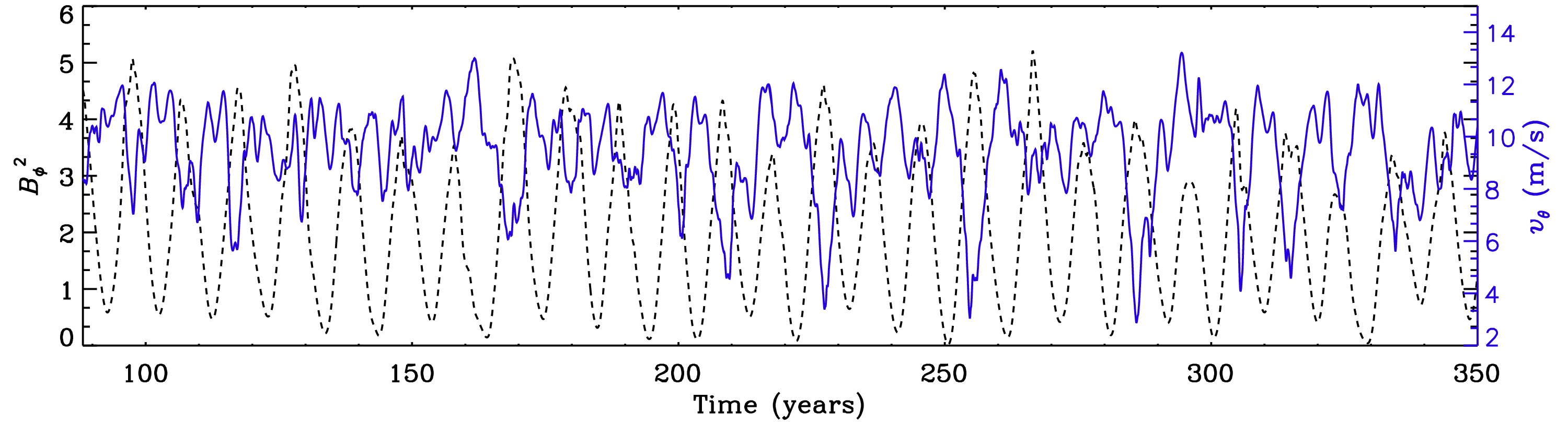}
\caption{Temporal variation of the net surface meridional flow 
$v_\theta$, averaged over 5--35$^\circ$ latitudes
(solid curve) with the solar cycle (dashed). The solar cycle is 
measured by the energy density of the toroidal field at $0.71R_\odot$ 
averaged over the same latitudes and normalized by $B_0^2$.
%, where $B_0 = 5 \times 10^4$~G. 
}
\label{fig:solcycle}
\end{figure*}

In \Fig{fig:bfly}(c-d), we show the magnetic butterfly diagrams of $B_r$ and $B_\phi$. 
Although these magnetic cycles are, in general, similar to the ones obtained from the kinematic counterpart, there are many changes.
Particularly, in the non-kinematic model, the magnetic field is more irregular; the field strength changes from cycle to cycle, as seen more clearly in 
\Fig{fig:solcycle} (dashed line, which is a measure of the toroidal field), 
obtained from longer data. 
While in the kinematic dynamo models, the variability in the solar cycle is produced by invoking additional mechanisms (e.g., fluctuations in the active region properties; \cite{KM17, Kumar24} and meridional flow; \citet{Kar10, Hazra26}), here the feedback of the magnetic field is enough to produce a noticeable variation, although these are still lower compared to the observed variation in the solar cycle \citep{Biswas23}.  Nonlinearity also produces a hemispheric asymmetry in magnetic fields as predicted in previous nonlinear models through the coupling of symmetric and antisymmetric modes \citep[e.g.,][]{Tob97, KTW98, WT16}. 
Similar to the kinematic model, the equatorward migration of the toroidal field
at the base of the CZ and the poleward migration of the surface radial field are maintained.
In general, the magnetic cycles from the nonkinematic case more closely resemble solar observations.

\begin{figure}
\includegraphics[scale=0.6]{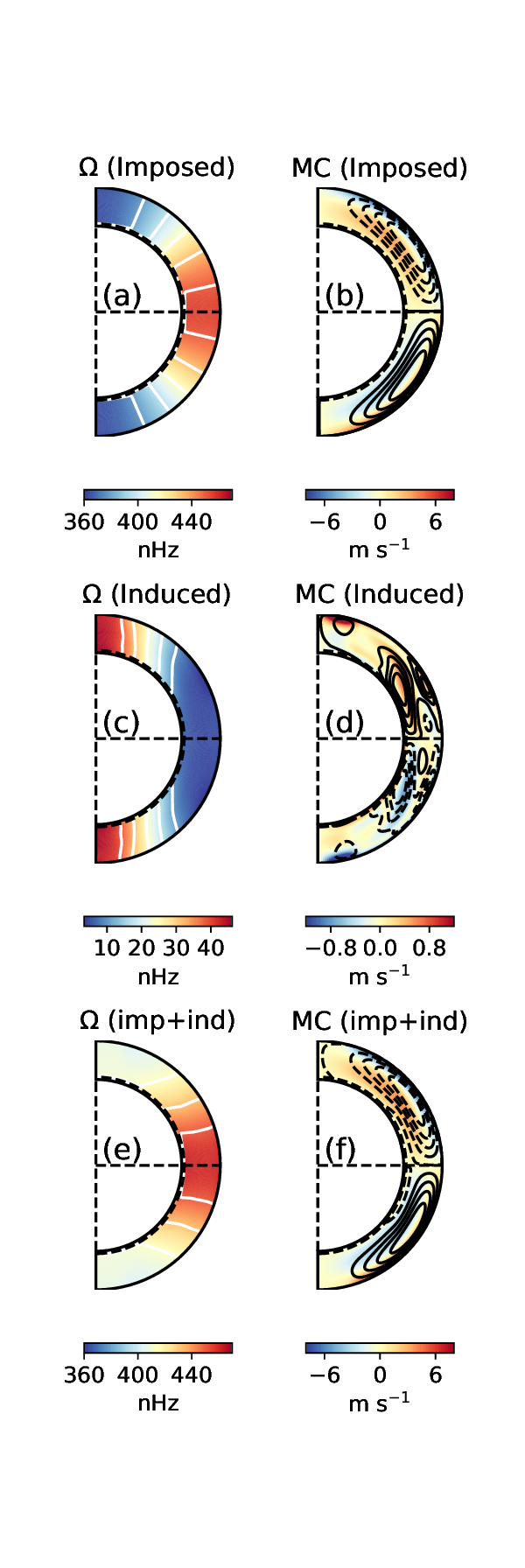}
\caption{Left panels: differential rotation $\Omega/2\pi$, right panels: 
meridional circulation; dashed (solid) curves correspond to anti-clockwise (clockwise) flow
and the colour represents the latitudinal flow speed $v_\theta$.
The top, middle, and bottom panels display the mean imposed flows, 
the induced flows obtained from the model, and the net flows (i.e., the sum of the mean and the induced flows acting on the magnetic induction equation), respectively.
}
\label{fig:meanflow}
\end{figure}

\subsection{Meridional circulation}
\label{sec:mc}
Following the pragmatic approach of previous flux transport dynamo models, in our model, a single-cell (in each hemisphere) meridional flow is also imposed; see \Fig{fig:meanflow}(b). This flow is poleward near the surface
with an amplitude of about 10~\mps and equatorward near the base of the CZ 
with an amplitude of about 2~\mps.  
The perturbed meridional circulation induced in our model
is shown in \Fig{fig:meanflow}(d). We note that to produce this plot, the meridional flow is 
%first averaged over longitudes and then 
averaged over a long time window
for which the butterfly diagram is shown in \Fig{fig:bfly}(c-d) i.e., for the data of 6 cycles.
Due to this induced meridional flow, the net flow (steady/axisymmetric $+$ perturbed) that is acting in the magnetic induction equation (\ref{eq:indy}) is changed from its usual steady meridional flow (\Fig{fig:meanflow}), and consequently the resulting magnetic cycle.
The induced \mc\ consists of multiple cells in the hemisphere and is mainly in the clockwise direction, thereby trying to oppose the imposed (background) flow.
%BBK: To work on the name.
The strength of the perturbed \mc, however, is much less compared to its steady part (about $10\%$)
and therefore the net flow is still capable of maintaining the equatorward transport of the toroidal flux at the base of the CZ.

The large-scale magnetic field that provides feedback on the \mc\ varies quasi-periodically in time due to the dynamo action.
We, therefore, observe a clear variation in the total \mc---becoming weaker during the solar maximum and stronger during the solar minimum. The amount of peak-to-peak variation of the \mf\ at the solar surface in the sunspot latitudes band is about $65\%$; see \Fig{fig:solcycle}.
%BBK: To give correlation coefficient.
This variation of \mf\ with the solar magnetic cycle is in good agreement with the surface observations of \citet{HR10}; see their Figure 4 \citep[also see][]{CD01, Beck02, Komm15}. 
We note that in observations of \citet{HR10}, there are data for only about two solar cycles
for which the variation of the surface \mf\ is about $55\%$.
In addition to the variation within the solar cycle, the \mc\ also varies cycle-to-cycle, which is again in agreement with the variations \citep{Ulrich10}.

\begin{figure*}
\includegraphics[scale=0.35]{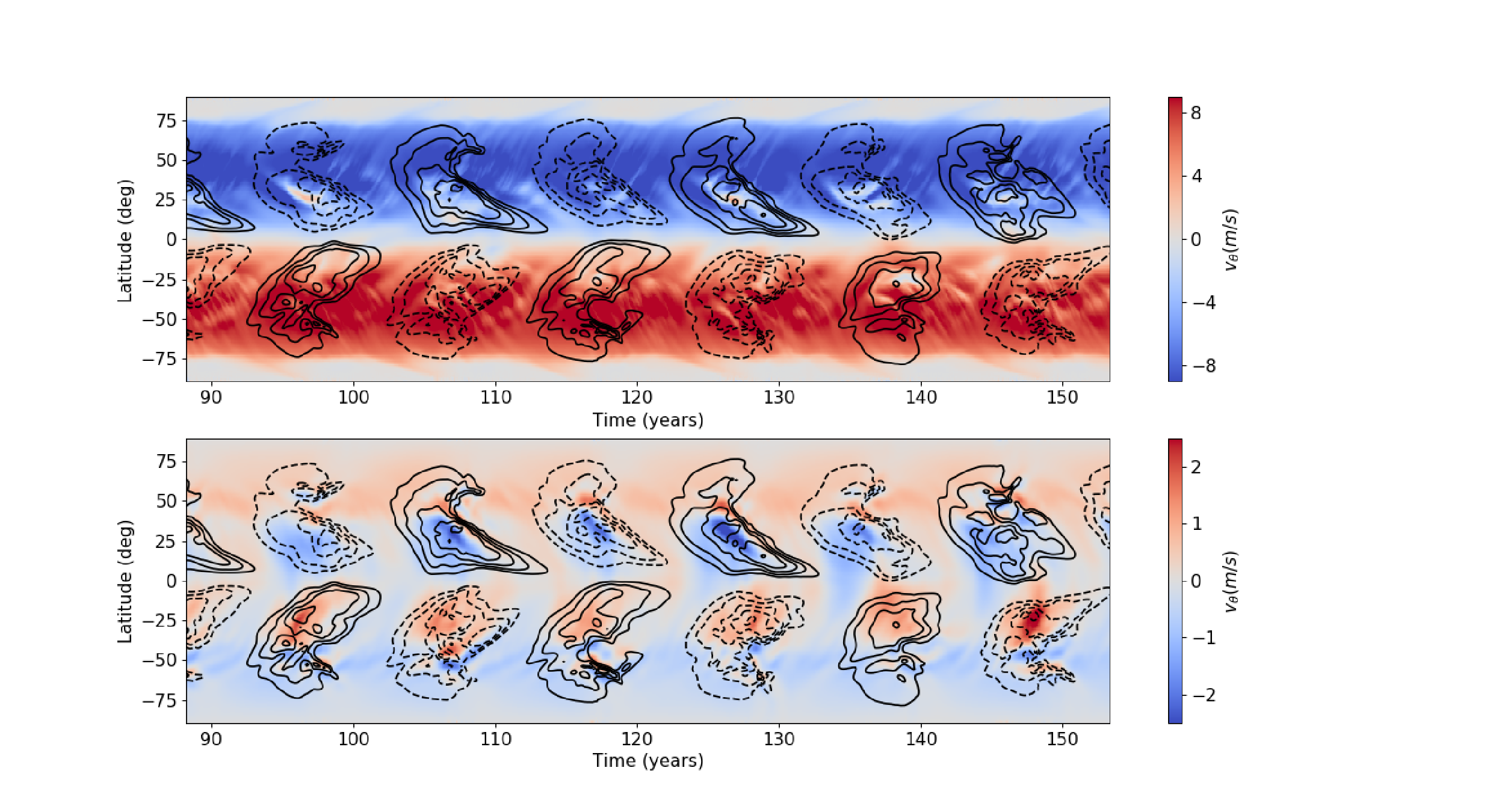}
\caption{Top and bottom panels represent the butterfly diagram of the meridional flow on the surface
($r=0.9997\Rs$) and bottom ($r=0.711\Rs$), respectively.
The overplotted contours mark the iso-contours of the strong toroidal field $B(\theta,0.711R_\odot)$.
}
\label{fig:bflyMC}
\end{figure*}

In the time-latitude diagram (\Fig{fig:bflyMC}), we clearly see that the \mf\ becomes weaker at
low latitudes exactly where the toroidal field is strong. In high latitudes (latitudes $> 75^\circ$) near the surface, we observe a prominent clockwise cell (\Fig{fig:meanflow}d), resulting in an equatorward flow for all the time.
This is somewhat consistent with the observation of \citet{Komm15, Dik10}, although the high latitudes data are less reliable and time-varying. We note that \citet{HC17} have shown
that the \mc\ becomes weaker during the solar maximum due to the generation of positive vorticity
(which produces clockwise \mc) through the large-scale Lorentz force. However, they did not consider the angular momentum balance, which is perturbed by magnetic tension (large-scale magnetic stress). It has been realised through the global convection simulations \citep{Guerr16, Passos17} and mean-field MHD models
\citep{Remp06} that the \mc\ in the CZ is primarily maintained by the redistribution of the angular momentum
through the zonal torque, so-called the Gyroscopic pumping \citep[][and references therein]{MH11}.
The zonal torque of the large-scale magnetic field (which is produced from the large-scale dynamo)
can significantly alter the meridional flow. 
%We guess that, had \citet{HC17} considered the angular momentum balance in their model, they would not have to reduce the magnetic forcing term arbitrarily to match the observations.

\begin{figure*}
\includegraphics[scale=0.35]{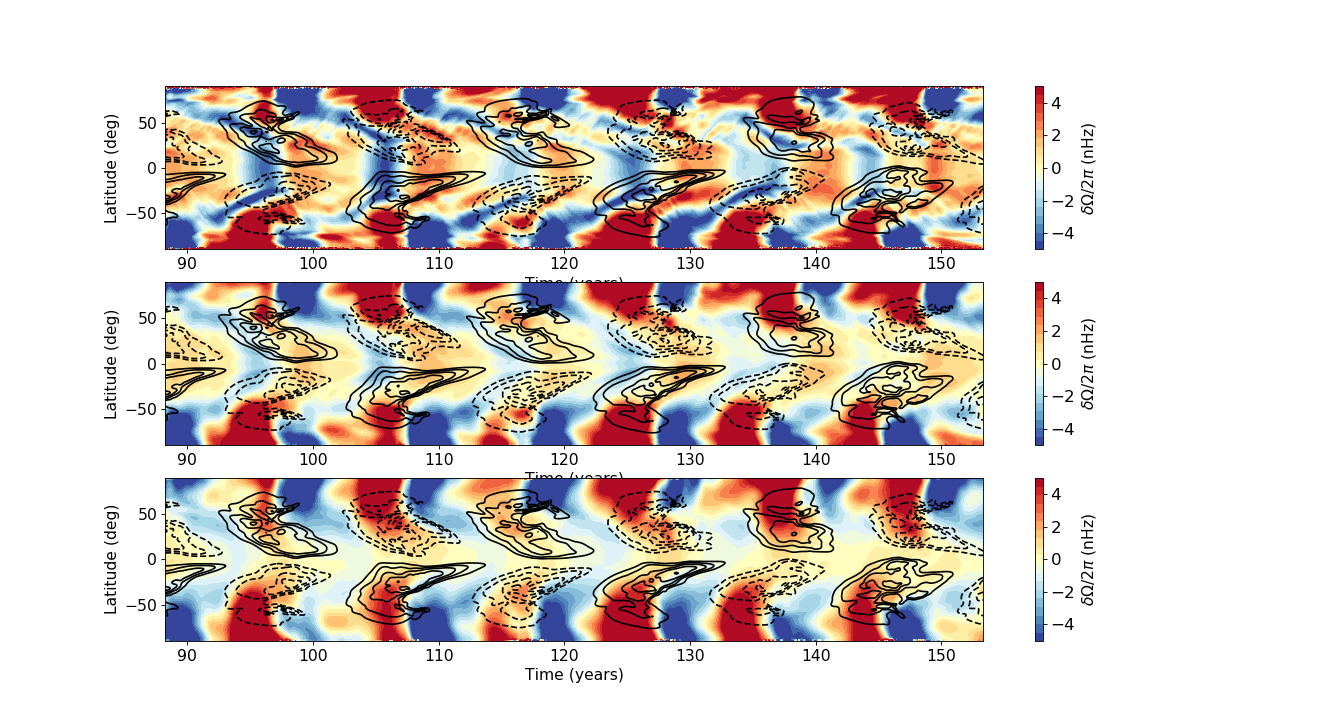}
\caption{Butterfly diagram of the torsional oscillation 
%$u_\phi(r,\theta,t) - \overline{u}_\phi(r,\theta)$ (where $\overline{u}_\phi(r,\theta)$ is the zonal and tomporal mean) 
$\delta\Omega_\phi(r,\theta,t)/2\pi$ 
at surface ($r=0.999\Rs$), mid-CZ ($r=0.85\Rs$) and base of the CZ ($r=0.71\Rs)$. 
The iso-contours represent the strong toroidal field $B_\phi(\theta,0.71\Rs)$.}
\label{fig:to}
\end{figure*}

\subsection{Differential rotation and torsional oscillation}
\label{sec:dr}
The imposed mean differential rotation, which is an approximate analytic fit to the helioseismic data, is shown in \Fig{fig:meanflow}(a).
The induced differential rotation as measured by 
%$\brac{\Omega(r,\theta)}_{\phi,t} = \brac{v_\phi(r,\theta,\phi,t)}_{\phi,t}/r\sin\theta$, averaged over longitude and 
$\brac{\Omega(r,\theta)}_{t} = \brac{v_\phi(r,\theta,t)}_{t}/r\sin\theta$, averaged over 
time (for six solar cycles), 
is shown in \Fig{fig:meanflow}(c). 
This induced rotation is faster at higher latitudes than at the equator. This is expected because the meridional circulation transports angular momentum poleward---thereby spinning up the polar region.
%as no Reynolds stress to transport equatorward angular momentum.
%In the absence of any equator-ward angular momentum transport, we expect the 
%polar region to spin up to conserve angular momentum.
The net differential rotation (induced + imposed) is therefore significantly changed, and the pole-equator contrast has reduced to about $10\%$ 
from its imposed value of about $25\%$ (\Fig{fig:meanflow}(a-e)). 
This, therefore, reduces the growth rate of the toroidal field generation.

%Changes in differential rotation $\brac{\Omega (r,\theta,t)}_\phi$ i.e., the torsional oscillation as measured by $\delta\Omega(r,\theta,t)/2\pi = [\brac{v_\phi(r,\theta,\phi,t)}_\phi - \brac{v_\phi(r,\theta,\phi,t)}_{\phi,t}]/(2\pi~r \sin\theta)$ is shown in \Fig{fig:to}.
Changes in differential rotation $\brac{\Omega (r,\theta,t)}_\phi$ i.e., the torsional
oscillation as measured by
$\delta\Omega(r,\theta,t)/2\pi = [v_\phi(r,\theta,t) - \brac{v_\phi(r,\theta,t)}_{t}]/(2\pi r \sin\theta)$
is shown in \Fig{fig:to}.
We find that a band rotating faster than average starts in high latitudes ($>50^\circ$)
and moves equatorward. 
The poleward branch of this faster-rotating band is, however, weak. On examining the observed data \cite[Figure 25 of][]{H09}, which shows a less prominent signature of this branch in the subsurface layers, we find that our results are not drastically different from observations. 
It is interesting to note that in the mean-field model of \citet{Remp06}, the poleward branch was robustly reproduced
as a consequence of Lorentz feedback on the flow, while to produce the equatorward branch, a thermal forcing was needed at the low latitudes. In our model, the low-latitude branch is a dominant feature and starts at relatively high latitudes.
The overall amplitude of the torsional oscillation is about $2\%$
while in observation it is about $1\%$ \cite[cf.\ Figure 25 of][]{H09}.
In agreement with observation, the \to\ begins a few years before the solar cycle starts (\Fig{fig:to}), forming the so-called extended solar cycle \citep{Srivastava18}.
Here, we recall that in our model, the solar cycle is identified by the strength of the toroidal field at the base of the CZ rather than by the emergence of active regions.

We observe a slight variation in \to\ with depth. As seen in \Fig{fig:to}, the overall amplitude and morphology remain the same at three depths. These are again in good agreement with observations, although the data in the deeper CZ are sparser and more intermittent.

\section{Conclusions and Discussion}
\label{sec:con}
We have presented our preliminary results of a non-kinematic flux transport dynamo
model.  While this model captures the back reaction of the dynamo-generated large-scale magnetic field in a similar way as done in many non-kinematic dynamo models and parameterises the poloidal field generation (the Babcock--Leighton process) through an axisymmetric $\alpha$
term near the surface, our model does not compute the mean flows; instead, they are imposed (guided by observations). This eliminates the need for free parameters to prescribe the $\Lambda$ effect. 
Our model is developed by utilising the 3D MHD code Rayleigh and by solving the perturbations of the observed mean flows (which act back on the dynamo equation) and the thermodynamics. 

The key findings of our study are the followings.
(i) The large-scale Lorentz force of the dynamo-generated magnetic field
induces a meridional circulation in such a way that the net flow is reduced near the
solar maximum and thereby produces a clear anti-correlation with the solar magnetic
cycle. In addition to the variation
within the cycle, meridional flow shows cycle-to-cycle variation.  All these findings are in good agreement with observations \citep{CD01, HR10, Ulrich10, Komm15}.
(ii) Our model also produces torsional oscillation with an amplitude slightly higher than the observed one. The equatorward migration of the faster and slower-rotating bands and their times of appearance are in good agreement with observations \citep{H09}.
(iii) The magnetic cycle with correct features (e.g., polarity reversal, 11-year periodicity, and equator-ward migration) persists.
%even with the inclusion of the perturbed flow in the mean (imposed) one. 
The magnetic field is more irregular and fluctuating, and the cycle displays considerable variation and weak hemispheric asymmetry. 

Above favourable results encourage us to implement the active region eruptions in this model \citep[as done in STABLE;][]{MT16, KM17} and explore the possible implications of non-axisymmetric magnetic field, such as active longitudes, the tachocline
instabilities, global magneto-shear
instabilities and the related phenomenon of magnetic
Rossby waves \citep{GF97, DG99, Cally03}. Performing simulations with this model under more solar-like parameters using helioseismic observations of flows \citep{Wavhal25} will also help identify the importance of nonlinearity in driving hemispheric asymmetry, long-term modulations, and grand minima \citep{BTW18, das22, SN25, Hazra26}.  We plan these for our future projects.

%%%%%%%%%%%%%%%%%%%%%%%%%%%%%%%%%%%%%%%%%%%%%%%%%%%%%%%%%%%%%%%%%%%%
\section*{Acknowledgements}
The Rayleigh code used in this study was developed by Nicholas Featherstone with support from the National Science Foundation through the Computational Infrastructure for Geodynamics (CIG). This effort was supported by NSF grants NSF-0949446 and NSF-1550901.
The authors also thank Nicholas for helping us code additional terms in Rayleigh and the anonymous reviewer for providing useful comments.  
%Computations were carried out with resources provided by NASA's High-End Computing program (Pleiades) when B.B.K. was a Jack Eddy Postdoctoral Fellow (supported by the NASA Living With Fellowship Program, administered by the University Corporation for Atmospheric Research).
Computations were carried out with resources provided by NASA's High-End Computing program (Pleiades) and NCAR (Yellowstone).
B.B.K. acknowledges the financial support from the Anusandhan National Research Foundation (ANRF) through the MATRIC program (file no. MTR/2023/000670). 
%%%%%%%%%%%%%%%%%%%%%%%%%%%%%%%%%%%%%%%%%%%%%%%%%%
%\section*{Data Availability}

\bibliographystyle{spr-mp-sola}
\bibliography{paper}
\label{lastpage}
\end{document}